\documentstyle[prd,aps,epsf,eqsecnum]{revtex}
\begin{document}

\title
{One-loop $\lambda \phi^4$ field theory in Robertson-Walker 
spacetimes: adiabatic regularization and analytic approximations}

\author
{Carmen Molina-Par\'{\i}s$^{+}$
\thanks{electronic address: carmen@t6-serv.lanl.gov},
Paul R. Anderson$^{++,+}$
\thanks{electronic address: anderson@wfu.edu},
and
Stephen A. Ramsey$^{+++}$
\thanks{electronic address: saramsey@u.washington.edu},
}

\address{
$^{+}$
Theoretical Division T-8, Los Alamos
National Laboratory, Los Alamos, New Mexico, 87545\\
$^{++}$
Department of Physics
Wake Forest University, Winston-Salem,
North Carolina, 27109\\
$^{+++}$
Genome Center,
University of Washington, Seattle, Washington, 98195-2145
}

\maketitle

\bigskip

PACS number(s): 04.62.+v, 11.10.Gh, 98.80.Cq

\bigskip

\abstract 

{The renormalization of a scalar field theory with a quartic
self-coupling (a $\lambda \phi^4$ theory) via adiabatic regularization
in a general Robertson-Walker spacetime is discussed.  The adiabatic
counterterms are presented in a way that is most conducive to
numerical computations.  A variation of the adiabatic regularization
method is presented which leads to analytic approximations for the
energy-momentum tensor of the field and the quantum contribution to
the effective mass of the mean field.  Conservation of the
energy-momentum tensor for the field is discussed and it is shown that
the part of the energy-momentum tensor which depends only on the mean
field is not conserved but the full renormalized energy-momentum
tensor is conserved as expected and required by the semiclassical
Einstein's equation. It is also shown that if the analytic
approximations are used then the resulting approximate energy-momentum
tensor is conserved.  This allows a self-consistent backreaction
calculation to be performed using the analytic approximations.  The
usefulness of the approximations is discussed.}

\section{Introduction}

The study of free quantized fields in curved space has been a
remarkably fruitful endeavor, particularly in the applications that
have been made to black hole and cosmological spacetimes~\cite{b-d}.
Much less has been done regarding interacting fields in curved space.
However, interacting fields are very important since all real fields
in nature appear to have interactions.  Interactions also play an
important role in cosmological models such as inflation, being
required in some cases for the inflaton potential to have the right
form and also for the thermalization that is necessary to reheat the
universe after inflation.  Interactions can also significantly enhance
the particle production that often occurs for free fields in curved
space.

The study of interacting quantum fields in Robertson-Walker (RW)
spacetimes is of great importance as well in understanding quantum
fields in Minkowski space.  It is well known that following a
relativistic heavy ion collision, the quark-gluon plasma produced
eventually undergoes a chiral phase transition. A good approximation
to describing the dynamics of this system is provided by the linear
$\sigma$ model, and by assuming that the expansion is mostly
radial~\cite{l-d-c,c-k-m-p}. Written in terms of the spherical
hydrodynamical fluid coordinates the system is equivalent to an
interacting quantum scalar field (mean field plus fluctuations) in a
RW spacetime~\cite{l-mp}.

Perhaps the simplest interacting quantum field theory in four
dimensions is a scalar field with a quartic self-coupling, often
called the ``$\lambda \phi^4$'' theory.  There is a long history of
the study of this theory in curved space.  The original investigations
centered on renormalization. Drummond~\cite{drummond}, Birrell and
Ford~\cite{b-ford1,b-ford2}, Bunch, Panangaden, and
Parker~\cite{b-p-p1}, and Bunch and Panangaden~\cite{b-p2}
investigated the renormalization of the theory in various cosmological
spacetimes using techniques such as dimensional regularization.  Bunch
and Parker~\cite{b-p3} showed that the theory is renormalizable in an
arbitrary spacetime to second order in the coupling constant
$\lambda$.  Birrell~\cite{birrell2} extended their work by using
momentum space techniques and computing self-energy graphs to second
order in $\lambda$.

Along with studies of the renormalization of the theory, various
calculations have been undertaken.  For example, Ford and Toms
investigated phase transitions caused by one-loop radiative
corrections in an expanding universe~\cite{f-toms}. The one-loop
finite temperature effective potential for a $\lambda \phi^4$ theory
in a RW universe was calculated by Hu, under the assumptions that the
rate of change of quantum fluctuations is much greater than that of
the mean field and the expansion rate of the universe~\cite{hu}.
Ringwald investigated the evolution of the expectation value of the
quantum fluctuation $\langle \psi^2 \rangle$ at one-loop order in a
spatially flat RW universe~\cite{ringwald}.

The quantity $\langle \psi^2 \rangle$ can be used to determine the
backreaction of the quantum fluctuations on the mean (or classical)
scalar field as it appears as an effective mass for the mean field at
one-loop order.  However, to determine the backreaction of the scalar
field on the spacetime geometry, one must compute the renormalized
energy-momentum tensor for the field.  The renormalization of the
energy-momentum tensor to one-loop order in a spatially flat RW
spacetime was discussed by Paz and Mazzitelli~\cite{m-paz1}.  They
displayed the renormalization counterterms which were obtained using
point splitting, adiabatic regularization, and dimensional
regularization.  The divergent counterterms were displayed in the
format of dimensional regularization.  Mazzitelli, Paz, and El Hasi
used this formalism in a calculation relating to the evolution of the
inflaton and the reheating after inflation in the new inflationary
scenario~\cite{m-paz-h}.

There are two ways that so called ``nonperturbative'' effects are
usually taken into account.  One is the Hartree approximation which
works for a single scalar field~\cite{hartree}.  The other is the
large $N$ approximation where a single scalar field is replaced with
$N$ scalar fields which are coupled via a quartic interaction, which
is invariant under the group $O(N)$ and is thus often called the
$O(N)$ model.  Mazzitelli and Paz considered the renormalization in
both the Hartree and large $N$ approximations in an arbitrary
background gravitational field, using point splitting techniques, and
adiabatic and dimensional regularization~\cite{m-paz2}.  More recent
work on the backreaction of a scalar field on the background geometry
in a RW spacetime has been done for inflationary models by Boyanovsky,
Cormier, de Vega, Holman, Kumar, Lee, Singh, and
Srednicki~\cite{boyan1,boyan2,boyan3} and by Ramsey and Hu~\cite{r-h}.

In this paper we derive a set of renormalized equations that can be
used to determine the evolution of a quartically coupled scalar field
with arbitrary mass and curvature coupling to one-loop order in a RW
spacetime.  We also derive expressions for the unique components of
the renormalized energy-momentum tensor that can be used to determine
the backreaction of the field on the spacetime geometry.  To
renormalize we use the method of adiabatic regularization which is
particularly useful for deriving a set of equations which are to be
solved numerically.  We display the counterterms for the
energy-momentum tensor and the quantum contribution to the effective
mass of the mean field (for arbitrary mass and curvature coupling).
Previously the adiabatic counterterms for the energy-momentum tensor
have been displayed by Paz and Mazzitelli~\cite{m-paz1} but only in
the context of dimensional regularization, which makes it difficult to
use them for numerical computations, and by Ramsey and Hu for the
minimally coupled case (if one takes the one-loop limit of their $1/N$
expansion)~\cite{r-h}.

We discuss in more detail than has previously been done the
conservation of the energy-momentum tensor for the full system (mean
field plus quantum fluctuations). We show that the natural division of
this tensor into a ``classical'' and a ``quantum'' piece leads to
neither piece being separately conserved.  We also show explicitly
that the full energy-momentum tensor is conserved.

We present a variation on the method of adiabatic regularization which
has been used by one of us to develop an analytic approximation for
the energy-momentum tensor for a free scalar field in a RW
spacetime~\cite{paul99}.  We use this method to derive analytic
approximations to both the energy-momentum tensor of the quantum
fluctuation and to the effective mass of the mean field.  If the
analytic approximation is used in the equation for the mean field and
if it is also used for the ``quantum'' energy-momentum tensor, then
the resulting set of equations results in a conserved approximate
energy-momentum tensor. Thus the analytic approximations can be used
in lieu of the full renormalized expressions in the mean field and
semiclassical backreaction equations. They are useful for the
investigation of vacuum polarization effects, but not particle
production since particle production is a nonlocal phenomenon.
Nevertheless, the approximations give important information,
particularly if one wishes to estimate the conditions under which the
loop expansion breaks down.  We discuss the validity of the
approximation and argue that it is likely to be most useful for
massless fields.

It is well known that the problem of solving Einstein's equation in
the presence of quantum matter is not an easy one.  On the
left-hand-side of Einstein's equation one needs higher order geometric
tensors (${}^{(1)}H_{\mu\nu}$, ${}^{(2)}H_{\mu\nu}$, and $H_{\mu\nu}$)
that involve up to fourth order time derivatives of the metric
$g_{\mu\nu}$~\cite{b-d}. On the right-hand-side one encounters the
expectation value of the energy-momentum tensor of the quantum field
in a certain quantum state (for which there is no a priori rule to be
determined or chosen), a quantity that is ultraviolet divergent. One
has to regularize and renormalize $\langle T_{\mu\nu}\rangle$ in such
a way that it remains covariantly conserved, as the left-hand-side of
Einstein's equation is. This is why the problem of a full backreaction
of the quantum matter on the spacetime geometry is so difficult. One
has to make use of regularization methods that are suited for a
numeric computation (as the dynamical equation for the quantum field
is most likely not to have analytic solutions), and at the same time
guarantee the covariant conservation of the renormalized value of
$\langle T_{\mu\nu}\rangle$.  In this paper we use adiabatic
subtraction to fulfill both requirements, and to set up all the
formalism and techniques required to perform a  backreaction
calculation for an interacting theory (mean field plus fluctuations)
in a general RW spacetime.

Before involving ourselves in analysing the backreaction, it will be
helpful to estimate how big or small the quantum effects on the
geometry and the mean field are. In this paper we introduce and
describe the analytic approximation as such a tool. It yields a
renormalized and covariantly conserved energy-momentum tensor for the
quantum fluctuations, (that carries no information, whatsoever, about
particle production effects), that can be evaluated in the spacetime
geometry that is a solution of the Einstein's equation determined by
the mean field.  This paper presents all the technical details for
such a calculation. Future work will consist of evaluating the
analytically approximated energy-momentum tensor of the quantum
fluctuations in various scenarios, such as the reheating period of the
inflationary regime of the early universe, and the spherical expansion
of the quark-gluon plasma.

In section \ref{sec:notation} we introduce the conventions to be used,
the background geometry, and derive the one-loop equations for the
mean and quantum fields. We also compute the energy-momentum tensor of
the system at one-loop, and show that it splits naturally in two
terms: a ``classical'' and a ``quantum'' energy-momentum tensor.  In
section \ref{sec:adiabatic} we discuss the method of adiabatic
regularization when the quantum fluctuations have a time dependent
mass, and derive the adiabatic order two and four counterterms that
need to be subtracted from $\langle \psi^2 \rangle_u$ and $\langle
T_{\mu \nu}\rangle_u$, respectively. We explicitly separate those new
terms that were not present in the free case~\cite{B,AP}. In section
\ref{sec:analytic} we introduce the analytic approximation as a way of
estimating the importance of vacuum polarization effects, and as a
first approximation to doing a full backreaction calculation. In
section \ref{sec:conservation} we discuss the covariant conservation
of the energy-momentum tensor.  We show that the part of the
energy-momentum tensor that depends only on the mean field is usually
not conserved by itself, but that both the full energy-momentum tensor
and its analytic approximation are covariantly conserved.

\section{Background, Conventions, and Notation}
\label{sec:notation}

We consider a quantum scalar field with self-interactions in a general
RW spacetime. The metric of an RW spacetime can be written in the
form\footnote{Throughout this paper we use units such that $\hbar = c
= 1$.  The metric signature is $(+ - - -)$ and the conventions for
curvature tensors are $R^\alpha_{\beta\gamma\delta} =
\Gamma^\alpha_{\beta\gamma,\delta} - ...$ and $ R_{\mu\nu} =
R^\alpha_{\mu\alpha\nu}$.}
\begin{equation}
{\mathrm{d}} s^2 
= a^2(\eta) \left[{\mathrm{d}} \eta^2 - \frac{{\mathrm{d}} r^2}
{1 - \kappa r^2} - r^2 {\mathrm{d}} \Omega^2 \right]
\label{eq:metric}
\; ,
\end{equation}
where $\eta$ is the conformal time coordinate, and $\kappa=-1,0,+1$ is
the three-dimensional spatial curvature, corresponding to spatial
Cauchy hypersurfaces that have negative, zero, and positive spatial
curvature, respectively.

The action of a scalar field with a quartic self-interaction is given
by
\begin{eqnarray}
S_{\rm matter}[\Phi , g_{\mu \nu}] &=& -\frac{1}{2} \int {\mathrm{d}}^4 x
\; 
{(-g)}^{\frac{1}{2}} \left[ \Phi (\Box +m^2 + \xi R) \Phi 
+ \frac{\lambda}{12}
\Phi^4
\right]
\label{eq:action}
\; ,
\end{eqnarray}
where $g$ is the determinant of the metric, $\Box$ the D'Alembert wave
operator given by $\Box = g^{\mu \nu} \nabla_\mu \nabla_\nu$, and $R$
the scalar curvature of the RW spacetime.

The equation of motion for the classical field (obtained by the
principle of least action) is given by
\begin{eqnarray}
 \left( \Box +m^2 + \xi R  + \frac{\lambda}{3!} \Phi^2
\right) \Phi&=& 0
\label{eq:eom}
\; .
\end{eqnarray}
The classical energy-momentum tensor is
\begin{eqnarray} 
T_{\mu \nu}    &=& 
(1 - 2\xi)
\partial_\mu  \Phi  \partial_\nu  \Phi
+ (2 \xi - 1/2) g_{\mu \nu}
 \partial_\alpha  \Phi  \partial^\alpha  \Phi  
-2 \xi \Phi  \nabla_\mu\nabla_\nu  \Phi  
\nonumber
\\
&+&
2 \xi g_{\mu \nu} \Phi  \Box  \Phi - \xi G_{\mu \nu} \Phi^2 
+\frac{m^2}{2}  g_{\mu \nu}  \Phi^2   
+\frac{\lambda}{4!}  g_{\mu \nu} \Phi^4  
\label{eq:tensorPhi}
\; . 
\end{eqnarray}
If we quantize the theory $\Phi$ becomes an operator.  We then define
the mean (or background) field $\phi$ by the equations
\begin{mathletters}
\begin{eqnarray}
 \Phi &\equiv&\phi +  \psi\; ,
\\
   \label{eq:meanfield} 
  \phi &\equiv& \langle \Phi \rangle
\; ,
\end{eqnarray}
\end{mathletters}
where the expectation value is taken with respect to the initial state
of the system (in the Heisenberg representation).

Taking the expectation value of Eq. (\ref{eq:eom}) and noting that
\begin{eqnarray}
\langle \Phi^3 \rangle&=&
\phi^3 + 3 \phi^2 \langle \psi \rangle + 3 \phi \langle \psi^2 \rangle
+\langle { \psi}^3 \rangle=
\phi^3  + 3 \phi \langle \psi^2 \rangle
+\langle { \psi}^3 \rangle
\; , 
\end{eqnarray}
we find the following equation for the mean field $\phi$
\begin{eqnarray}
 ( \Box +m^2 + \xi R ) \phi + \frac{\lambda}{3!}
\left( \phi^3  + 3 \phi \langle \psi^2 \rangle
+\langle { \psi}^3 \rangle 
\right) &=& 0
\label{eq:equationphi1}
\; .
\end{eqnarray}
In the same way, by subtracting equation (\ref{eq:equationphi1}) from
equation (\ref{eq:eom}), we obtain the equation of motion for the
quantum fluctuation $\psi$
\begin{eqnarray}
 ( \Box +m^2 + \xi R ) \psi + \frac{\lambda}{3!} 
\left( 3 \phi^2 \psi - 3 \phi
\langle  \psi^2 \rangle
+3 \phi \psi^2 + \psi^3 - \langle \psi^3 \rangle
\right) &=& 0
\label{eq:equationpsi1}
\; .
\end{eqnarray}
If we truncate at one-loop (free field theory for the quantum
fluctuation $\psi$), the equations of motion (\ref{eq:equationphi1})
and (\ref{eq:equationpsi1}) become~\cite{m-paz1}
\begin{mathletters}
\begin{eqnarray}
 ( \Box +m^2 + \xi R ) \phi + \frac{\lambda}{3!} \phi^3
+\frac{\lambda}{2!} \phi \langle \psi^2 \rangle
&=& 0
\; ,
\label{eq:equationphi2}
\\
 ( \Box +m^2 + \xi R ) \psi + \frac{\lambda}{2!} 
\phi^2 \psi 
 &=& 0 
\label{eq:equationpsi2} 
\; .  
\end{eqnarray} 
\end{mathletters}
The expectation value of the energy-momentum tensor can be broken into
a ``classical'' and a ``quantum'' part.  The classical part is given
by
\begin{mathletters}
\begin{eqnarray} 
\langle T_{\mu \nu} \rangle^{C} 
   &\equiv& 
(1 - 2\xi) \partial_\mu  \phi  \partial_\nu  \phi
+ (2 \xi - 1/2) g_{\mu \nu}
 \partial_\alpha  \phi  \partial^\alpha  \phi  
-2 \xi \phi  \nabla_\mu\nabla_\nu  \phi  
\nonumber
\\
&+&
2 \xi g_{\mu \nu} \phi  \Box  \phi - \xi G_{\mu \nu} \phi^2 
+\frac{m^2}{2}  g_{\mu \nu}  \phi^2   
+\frac{\lambda}{4!}  g_{\mu \nu} \phi^4  
\label{eq:tensorC} 
\; ,
\end{eqnarray}
while the quantum part is
\begin{eqnarray}
\langle T_{\mu \nu}\rangle^Q=\langle T_{\mu \nu}\rangle_u
   &\equiv&  
(1 - 2\xi) \langle \partial_\mu  \psi  \partial_\nu \psi  \rangle
+ (2 \xi - 1/2) g_{\mu \nu}
\langle \partial_\alpha \psi  \partial^\alpha  \psi  \rangle
-2 \xi \langle  \psi  \nabla_\mu\nabla_\nu \psi  \rangle
\nonumber
\\
&+&
2 \xi g_{\mu \nu} \langle \psi  \Box \psi  \rangle
- \xi G_{\mu \nu} \langle  \psi^2   \rangle
+\frac{m^2}{2}  g_{\mu \nu} \langle  \psi^2   \rangle
+\frac{\lambda}{4}  g_{\mu \nu}  \phi^2 \langle  \psi^2  \rangle
\; .
\label{eq:tensorQ}
\end{eqnarray}
\end{mathletters}
Equations (\ref{eq:equationphi2}--\ref{eq:tensorQ}) describe our
system at one-loop order (mean field $\phi$ plus quantum fluctuations
$\psi$, which contribute to the effective mass of $\phi$). It is well
known~\cite{collins,B} that to make sense of these equations one needs
to regularize the theory, that is, on the one hand, define a way to
obtain from the bare parameters ($m,\lambda$, and $\xi$) the
renormalized ones, and on the other hand, regularize the divergent
quantities $\langle \psi^2 \rangle_u$ and $\langle T_{\mu
\nu}\rangle_u$, to obtain the physically finite energy-momentum tensor
of the system.  In the next section we discuss these issues.

\section{Adiabatic Regularization} 
\label{sec:adiabatic}

Using dimensional regularization it has been shown that for a $\lambda
\phi^4$ theory in a general spacetime, the bare and the renormalized
parameters are related in the following way~\cite{b-p-p1,b-p2}
\begin{eqnarray} 
m^2_B  &\equiv& m^2_R - \frac{3 \lambda_R}{8 \pi^2 (n-4)}m^2_R
\; ,
\nonumber
\\
\xi_B - \frac{1}{6} &\equiv& \xi_R-\frac{1}{6} 
-\frac{3 \lambda_R}{8 \pi^2 (n-4)} \left( \xi_R - \frac{1}{6}\right)
\; , 
\nonumber\\
\lambda_B  &\equiv& \lambda_R - \frac{9 \lambda_R^3}
{8 \pi^2 (n-4)}
\; .
\end{eqnarray}
Here $n$ is the number of dimensions the spacetime has been
analytically continued to.  Thus even for a general spacetime the
counterterms for the renormalization of $m^2$, $\xi$, and $\lambda$ are
constant in space and time as expected.

In principle one would prefer to renormalize at the level of the
effective action and then vary that action with respect to $\phi$ and
the metric $g_{\mu\nu}$ to obtain the renormalized equations of motion
and energy-momentum tensor, respectively~\cite{guido}.  However the
computation of the one-loop effective action for arbitrary mean fields
$\phi$ and arbitrary spacetime geometries is quite involved, and will
not yield an intrinsically different answer from that obtained by
looking at the one-loop field equations.  For this reason, we consider
a different method called adiabatic regularization~\cite{P,PF,FP,FPH}
which works at the level of the field equations. Another advantage of
adiabatic regularization is that it is well suited to perform
numerical calculations~\cite{A34,SA}.  In adiabatic regularization the
divergences in quantities such as $\langle \psi^2 \rangle_u$ and
$\langle T_{\mu\nu} \rangle_u$ are computed using a WKB expansion for
the modes of the quantized field $\psi$.  These terms are then
subtracted from the unrenormalized (bare) expressions with the result
that
\begin{mathletters}
\begin{eqnarray}
{\langle \psi^2 \rangle}_R &=& \langle \psi^2 \rangle_u 
- \langle \psi^2 \rangle_{ad} \; ,
 \label{eq:psiad} \\
 \langle T_{\mu\nu} \rangle_R &=& \langle T_{\mu\nu} \rangle_u 
- \langle T_{\mu\nu} \rangle_{ad}
\; , \label{eq:tmnad}
\end{eqnarray} 
where the subscripts $u$ and $ad$ stand for the unrenormalized (bare)
and the adiabatic value, respectively of ${\langle \psi^2 \rangle}$
and $ \langle T_{\mu\nu} \rangle$.  This procedure has been shown to
be equivalent to point splitting for free scalar fields in a RW
spacetime~\cite{Birrell,AP}.  For the quartically coupled scalar field
Eqs. (\ref{eq:equationphi2}) and (\ref{eq:equationpsi2}) then become
\begin{eqnarray}
 ( \Box +m^2_R + \xi_R R ) \phi + \frac{\lambda_R}{3!} \phi^3
+\frac{\lambda_R}{2!} \phi {\langle \psi^2 \rangle}_R  &=& 0\; ,  \\
 ( \Box +m^2_R + \xi_R R ) \psi 
+ \frac{\lambda_R}{2!} \phi^2 \psi &=& 0 \; .
\end{eqnarray} 
\end{mathletters}
In what follows we only consider the renormalized values of the
coupling constants $m$, $\xi$, and $\lambda$, so we drop the subscript
$R$ for these quantities.

We assume that the mean field is homogeneous $\phi=\phi (\eta)$, as a
RW spacetime is homogeneous and isotropic.  Then
Eq. (\ref{eq:equationphi2}) becomes
\begin{mathletters}
\begin{eqnarray}
\phi'' +  2 \frac{a'}{a} \phi'
+ a^2 \left( m^2_B + \xi_B R  + \frac{\lambda_B}{3!} \phi^2 
+\frac{\lambda_B}{2!} \langle \psi^2 \rangle \right)  \phi
&=&0
\; . \label{eq:phi-b}
\end{eqnarray} 
Here primes denote derivatives with respect to $\eta$.  The $\eta\eta$
component of the classical renormalized energy-momentum tensor
(\ref{eq:tensorC}) is given by
\begin{eqnarray} 
\langle T_{\eta \eta}\rangle^C_R&=& 
 \frac{1}{2} \phi'  \phi'
+ 6 \xi \frac{a'}{a} \phi \phi'
+3 \xi \left( \frac{a'^{\, 2}}{a^2} + \kappa \right) \phi^2
+\frac{a^2 m^2}{2}\phi^2
+\frac{a^2 \lambda}{4!}  \phi^4 
\; ,
\label{eq:t00-b}
\end{eqnarray}
and the trace is
\begin{eqnarray} 
\langle T \rangle^C_R &=& 
(6\xi-1)\frac{1}{a^2}
\phi'  \phi'
+ \frac{6}{a^2} \xi  \phi  \phi''
+\frac{12}{a^2} \xi \frac{a'}{a} \phi \phi'
+ \xi R \phi^2
+ 2 {m^2}\phi^2
+\frac{\lambda}{3!}  \phi^4 
\label{eq:t-b}
\; .
\end{eqnarray}
\end{mathletters}

To determine the renormalization counterterms used in adiabatic
regularization, (${\langle \psi^2 \rangle_{ad}}, \langle T_{\mu\nu}
\rangle_{ad}$), we first review canonical quantization in a RW
spacetime. We then discuss the WBK expansion for the modes of the
quantum fluctuation field $\psi$, and compute the adiabatic
counterterms needed to renormalize $\langle \psi^2 \rangle_u$ and
$\langle T_{\mu\nu} \rangle_u$.

Since at the one-loop level the quantum field $\psi$ is a free field
with an effective mass of the form $m^2 + \frac{\lambda
\phi^2}{2}$, it can be expanded in the following manner~\cite{b-d}
\begin{equation}
\psi(x) = \frac{1}{a(\eta)} \int {\mathrm{d}} \tilde{\mu}({\bf k}) 
\left[ a_{\bf k} \, Y_{\bf k}({\bf x}) 
\, f_k (\eta)
          + a_{\bf k}^\dagger \, Y_{\bf k}^*({\bf x}) \, 
f^*_k (\eta) \right]
\; ,
\end{equation}
where the measure is given by~\cite{B}
\begin{eqnarray}
\begin{array}{lllll}
  \int {\mathrm{d}} \tilde{\mu}({\bf k}) &\equiv& 
\int {\mathrm{d}}^3 {\bf k} \;\; & {\rm if}& \kappa = 0 \; , 
\vspace{0.2cm}
 \nonumber \\
   &\equiv& \int_0^{+ \infty} {\mathrm{d}} k \sum_{l,m} \;\; & {\rm if}&
\kappa = -1 \; ,
\vspace{0.2cm}
\nonumber \\
   &\equiv& \sum_{k,l,m} \;\; & {\rm if} & \kappa = +1 \; . 
\end{array}
\end{eqnarray}
The spatial part of the mode function, $Y_{\bf k}({\bf x})$,
obeys the equation
\begin{equation}
\Delta^{(3)} Y_{\bf k}({\bf x}) = - (k^2 - \kappa) Y_{\bf k}({\bf x})
\; ,
\end{equation}
and the time dependent part $f_k (\eta)$ is a solution to
the equation
\begin{eqnarray}
f''_k +  \omega_k^2 + \frac{\lambda }{2} a^2 \phi^2+ (\xi - 1/6) a^2 R &=& 0
\; .
\label{eq:modef}
\end{eqnarray}
Here $\omega_k^2 \equiv k^2 + m^2 a^2 $, primes denote derivatives
with respect to the conformal time $\eta$, and for spacetimes with the
metric (\ref{eq:metric}) the scalar curvature is given by
\begin{equation}
R = 6 \left(\frac{a''}{a^3} + \frac{\kappa}{a^2} \right) \;.
\end{equation}

The unrenormalized (bare) expression for the quantum part of the
energy-momentum tensor (\ref{eq:tensorQ}) is~\cite{B,AP}
\begin{mathletters}
\begin{eqnarray}
\langle {T^\eta}_\eta \rangle_{u} 
&=& \frac{1}{4 \pi^2 a^4}\int {\mathrm{d}}\mu(k) \left\{
|f_k'|^2 
+ \left( k^2+m^2 a^2+\frac{\lambda}{2} a^2 \phi^2\right) |f_k|^2 
             \right. \nonumber \\
      & & \left. \;\;\;\; + 6 \,\left(\xi - \frac{1}{6} \right) 
\left[\frac{a'}{a} (f_k {f^*_k}' 
    + f^*_k f_k') - \left(\frac{{a'}^{\,2}}{a^2} - \kappa
 \right) |f_k|^2 \right] 
  \right\} \; , \label{eq:t00u} \\
\langle T \rangle_{u} &=& 
\frac{1}{2 \pi^2 a^4}\int {\mathrm{d}}\mu(k) \left\{ \left(
m^2 a^2       + \frac{\lambda}{2} a^2 \phi^2 \right) |f_k|^2
+ 6 \left(\xi - \frac{1}{6}\right) \left[|f_k'|^2 
      - \frac{a'}{a} (f_k {f^*_k}' + f^*_k f_k') \right. \right. \nonumber \\
  &  &  \left. \left.   - \left(k^2+m^2 a^2 +\frac{\lambda}{2}a^2 \phi^2 
   + \frac{a''}{a} - \frac{a'^{\,2}}{a^2} 
           + \left(\xi-\frac{1}{6}\right) a^2 R \right) |f_k|^2 
\right]\right\}
\; .
\label{eq:tru}
\end{eqnarray}
Note that the pressure is not an independent quantity, and can be
obtained from knowledge of the energy density and the trace.

The equation for the mean field (\ref{eq:equationphi2}) also contains
the quantity $\langle \psi^2 \rangle$.  The unrenormalized expression
for it can be written in terms of the mode functions as follows
\begin{equation}
\langle \psi^2 \rangle_u = \frac{1}{2 \pi^2 a^2}\int 
{\mathrm{d}}\mu(k) \; |f_k|^2 
\; .
\end{equation}
\end{mathletters}
In these expressions the measure is 
\begin{eqnarray*}
\begin{array}{lllll}
 \int {\mathrm{d}} \mu(k) &\equiv& \int_0^{+\infty}
 {\mathrm{d}} k \;  k^2 \;\;\;
& {\rm if}& \kappa = 0, -1 \; ,
\vspace{0.2cm}
\nonumber \\
               &\equiv& \sum_{k=1}^{+ \infty} k^2 \;\;\; &{\rm if}&
\kappa = +1 \; .
\end{array}
\end{eqnarray*}

To determine the adiabatic counterterms needed to renormalize these
expectation values we solve the mode equation (\ref{eq:modef}) using a
WKB expansion.  To obtain this expansion we first make the variable
transformation
\begin{equation}
f_k = (2 W_k)^{-1/2} \exp\left[
\int^\eta {\mathrm{d}}\eta' \; W_k(\eta') \right]
\; .
 \label{eq:Wdef}
\end{equation}
Substituting Eq. (\ref{eq:Wdef}) into Eq. (\ref{eq:modef}) yields
\begin{equation}
W_k^2 = \omega_k^2 + \frac{\lambda}{2} a^2 \phi^2 
+ \left(\xi - \frac{1}{6}
 \right) a^2 R - \frac{1}{2} \left( \frac{W_k''}{W_k} 
- \frac{3}{2} \frac{W_k'^{\,2}}
 {W^2_k}\right)
\; . \label{eq:Weq}
\end{equation}
This equation is then solved iteratively with $\omega_k$ being of
adiabatic order zero and the next two terms on the right hand side
being of adiabatic order two\footnote{The $\lambda a^2 \phi^2$ term is
considered to be of second adiabatic order because only terms with up
to two time derivatives of $\phi$ are needed to cancel divergences in
$\langle T_{\mu\nu}\rangle_u$~\cite{m-paz1}.}. Thus to second
adiabatic order the solution for $W_k$ is
\begin{equation}
W_k = \omega_k + \frac{1}{2\omega_k^2} \left[ \frac{\lambda}{2} a^2  \phi^2
 + \left(\xi - \frac{1}{6}\right)  a^2 R 
- \frac{1}{2} \left( \frac{\omega_k''}
{\omega_k} - \frac{3}{2} \frac{\omega_k'^{\,2}} {\omega_k^2}\right)\right]
\; .
\label{eq:W2eq}
\end{equation}
To renormalize $\langle \psi^2 \rangle_u$ we use a second order
adiabatic expansion for the modes.  A fourth order expansion is
necessary to cancel the divergences in $\langle T_{\mu\nu} \rangle_u$
~\cite{B,AP}. The renormalization counterterms for these quantities
are
\begin{mathletters}
\begin{eqnarray}
{\langle \psi^2 \rangle}_{ad}&=& \langle \psi^2 \rangle^F_{ad} - 
\frac{1}{4 \pi^2 a^2} 
\int {\mathrm{d}} \mu (k)\; \frac{\lambda \phi^2 a^2}{4 \omega_k^3} \; ,\\ 
\label{eq:psi-ad}
\langle {T^{\eta}}_{\eta} \rangle_{ad}
&=& 
\langle {T^{\eta}}_{\eta}\rangle^{F}_{ad}
+
\frac{1}{4 \pi^2 a^4} \int {\mathrm{d}} \mu (k)
\left\{
\frac{\lambda \phi^2 a^2}{4\omega_k}
-\frac{\lambda^2 \phi^4 a^4}{32\omega_k^3}
+ \frac{m^2  \lambda( \phi^2 a^2 a'^{\, 2}  + \phi \phi' a^3 a')}{8\omega_k^5}
- \frac{5 m^4 \lambda \phi^2  a^4 a'^{\, 2} }{32\omega_k^7}
\right.
\nonumber
\\
& &  + \;  6\left( \xi -\frac{1}{6}\right)
\left.
\left[
-\frac{\lambda ( \phi^2 a'^{\, 2} 
  + 2 \phi \phi' a a' + \kappa \phi^2 a^2)}{8\omega_k^3}
+\frac{3 m^2 \lambda \phi^2 a^2 a'^{\, 2} }{8\omega_k^5}
\right]
\right\}
\; ,  
\label{eq:t00-ad}
\\
\langle T \rangle_{ad}
&=& 
\langle T \rangle^{F}_{ad}
+
\frac{1}{4 \pi^2 a^4} \int {\mathrm{d}} \mu (k)
\left\{
\frac{\lambda \phi^2 a^2}{2\omega_k} 
- \frac{\lambda \phi^2 a^4  (2 m^2+\lambda \phi^2)}{8\omega_k^3}
+\frac{m^2 \lambda}{32\omega_k^5} (3 \lambda \phi^4 a^6 
+ 4  a^4  \phi \phi'' +  4 a^4 \phi'^{\,2} \right. \nonumber \\
 & & \left. + \; 8 \phi^2  a^2 a'^{\, 2}   
 +8 \phi^2  a^3 a'' + 16 \phi \phi'  a^3 a')
 - \frac{5m^4 \lambda}{16\omega_k^7}
(4 \phi^2 a^4 a'^{\, 2}   +  \phi^2 a^5 a''  + 2 \phi \phi' a^5 a')
+ \frac{35m^6 \lambda a^6 a'^{\,2}}{32\omega_k^9} \right. 
\nonumber \\
& &  
\left.
 + \; 6\left(\xi -\frac{1}{6}\right)
\left[
- \frac{\lambda}{4\omega_k^3}(2 \phi \phi' a a'+ \phi^2 a a'' 
  + \phi \phi'' a^2 + \phi'^{\, 2} 
a^2 + \kappa \phi^2 a^2) 
\right. \right. \nonumber \\
& & \left. \left. + \; \frac{3 m^2 \lambda}{8\omega_k^5}
( 3 \phi^2 a^2 a'^{\, 2}  + 4 \phi \phi' a^3 a' 
+2  \phi^2 a^3 a'' + \kappa \phi^2 a^4)
-\frac{15 m^4 \lambda \phi^2 a^4 a'^{\, 2} }{8\omega_k^7}
\right]
\right\}
\; .
\label{eq:t-ad}
\end{eqnarray}
The adiabatic counterterms for the free field are~\cite{B,AP}
\begin{eqnarray} 
\langle \psi^2 \rangle^F_{ad} &=&
\frac{1}{4 \pi^2 a^2} \int {\mathrm{d}} \mu (k)
\left[ \frac{1}{\omega_k} 
- \left(\xi -\frac{1}{6}\right) \frac{a^2R}{2\omega^3_k} 
+\frac{m^2}{4\omega^5_k} (a'^{\, 2} +aa'')
-  \frac{5m^4a^2 a'^{\, 2} }{8\omega^7_k}
\right]  
\; ,
\label{eq:psi-ap}
\\
\langle {T^{\eta}}_{\eta}\rangle^{F}_{ad}
&=&
\frac{1}{4 \pi^2 a^4} \int {\mathrm{d}} \mu (k)
\left\{ 
\omega_k
+\frac{m^4a^2a'^{\, 2} }{8\omega_k^5}
-\frac{m^4}{32\omega_k^7}(2 a^2 a' a'''- a^2 a''^{\, 2} 
 + 4 a a'^{\, 2}  a'' - a'^{\, 4} )
+\frac{7m^6a^2}{16\omega_k^9}(a a'^{\, 2}  a'' + a'^{\, 4} ) 
\right. 
\nonumber \\
 & & 
- \; 
\frac{105m^8a^4a'^{\, 4} }{128\omega_k^{11}}
+ \left(\xi -\frac{1}{6}\right)
\left[
-\frac{3}{\omega_k}\left(\frac{a'^{\, 2} }{a^2}-\kappa  \right)
-\frac{3m^2a'^{\, 2} }{\omega_k^3}
+\frac{3m^2}{4\omega_k^5}\left( 2 a' a''' - a''^{\, 2} 
 - \frac{a'^{\, 4} }{a^2} \right) 
\right. \nonumber \\
& &  \left. - \; 
\frac{15m^4}{8\omega_k^7}\left(4 a a'^{\, 2} 
 a'' + 3 a'^{\, 4}  + \kappa a^2 a'^{\, 2}   
\right)
+\frac{105m^6a^2a'^{\, 4} }{8\omega_k^9}
\right] 
\nonumber\\
& &
\left. + \; \left( \xi -\frac{1}{6}\right)^2
\left[
-\frac{9}{2\omega_k^3}\left(\frac{2a'a'''}{a^2}
-\frac{a''^{\, 2} }{a^2}
-\frac{4a'^{\, 2} a''}{a^3}
-\frac{2 \kappa a'^{\, 2} }{a^2}
+\kappa^2  \right)
+\frac{27m^2}{\omega_k^5}\left(\frac{a'^{\, 2} a''}{a} 
+\kappa a'^{\, 2} \right)
\right]
\right\}
\; ,
\label{eq:t00-ap}
\\
\langle  T \rangle^{F}_{ad}
&=&
\frac{1}{4 \pi^2 a^4} \int {\mathrm{d}} \mu (k)
\left\{
\frac{m^2a^2}{\omega_k}
+\frac{m^4a^2}{4\omega_k^5}(a a'' + a'^{\, 2} )
-\frac{5m^6a^4a'^{\, 2} }{8\omega_k^7}
-\frac{m^4a^2}{16\omega_k^7}(aa'''' + 4 a' a''' + 3 a''^{\, 2}  )
\right.
\nonumber
\\
&&+ \; \frac{7m^6a^2}{32\omega_k^9}(4 a^2 a' a'''
+18 a a'^{\, 2}  a'' + 3 a^2 a''^{\, 2} +3 a'^{\, 4} )
-\frac{231m^8a^4}{32\omega_k^{11}}(aa'^{\, 2} a''+a'^{\, 4} )
+ \frac{1155m^{10}a^6a'^{\, 4} }{128 \omega_k^{13}}
\nonumber
\\
& &
+ \; \left(\xi -\frac{1}{6}\right)
\left[
-\frac{6}{\omega_k}\left(\frac{a''}{a}-\frac{a'^{\, 2} }{a^2}  \right)
-\frac{3m^2}{\omega_k^3}\left({2aa''}-a'^{\, 2}   +\kappa a^2 \right)
+\frac{9m^4a^2a'^{\, 2} }{\omega_k^5}
+\frac{3m^2}{2\omega_k^5}\left(aa''''- 2 \frac{a'^{\, 2}  a'''}{a} 
+\frac{a'^{\, 4} }{a^2} \right)
\right.
\nonumber
\\
& &
\left. - \; 
\frac{15m^4}{4\omega_k^7}\left(4 a^2 a' a''' + 3 a^2 a''^{\, 2} 
 +8 a a'^{\, 2}  a''
-a'^{\, 4}  +\kappa a^3 a'' +\kappa a^2a'^{\, 2}   \right)
  \right. \nonumber \\
& & \left.  + \; \frac{105m^6a^2}{8\omega_k^9}(8aa'^{\, 2}  a'' 
+ 5 a'^{\, 4}  + \kappa a^2a'^{\, 2} )
- \frac{945m^8 a^4 a'^{\, 4} }{8 \omega_k^{11}}
\right]
\nonumber
\\
& &
+ \; \left(\xi -\frac{1}{6}\right)^2
\left[
-\frac{9}{\omega_k^3}\left(
\frac{a''''}{a}
-\frac{4a'a'''}{a^2}
-\frac{3a''^{\, 2} }{a^2}
+\frac{6a'^{\, 2} a''}{a^3}
-\frac{2 \kappa a''}{a}
+\frac{2\kappa a'^{\, 2} }{a^2}  \right)
\right.
\nonumber
\\
&&
\left.
\left.
+ \; \frac{27m^2}{2\omega_k^5}\left(
{4a'a'''}
+{3a''^{\, 2} }
-{6\frac{a'^{\, 2} a''}{a}}
+{4\kappa a a''}
-{2\kappa a'^{\, 2} }
+a^2 \kappa^2  \right)
-\frac{135m^4}{\omega_k^7}\left({aa'^{\, 2} a''}
 +\kappa a^2 a'^{\, 2} \right)
\right]
\right\}
\; .
\label{eq:t-ap}
\end{eqnarray}
\end{mathletters}
Equations (\ref{eq:psi-ad}), (\ref{eq:t00-ad}), and (\ref{eq:t-ad})
[together with (\ref{eq:psi-ap}), (\ref{eq:t00-ap}), and
(\ref{eq:t-ap})] give the adiabatic counterterms needed to obtain the
renormalized expectation value of the quantum fluctuations and of the
quantum energy-momentum tensor for an interacting quantum scalar field
in a general RW spacetime, for the case of a homogeneous mean field
$\phi (\eta)$\footnote{Mazzitelli and Paz presented these counterterms
with the points separated and with the divergence structure given in
the context of dimensional regularization~\cite{m-paz1}.  Ramsey and
Hu carried out the adiabatic expansion up to order four for $\xi_R=0$
in the context of the leading $1/N$ expansion~\cite{r-h}.}.

\section{Analytic approximations}
\label{sec:analytic}

As it stands the method of adiabatic regularization can be used to
compute the quantities $\langle \psi^2 \rangle_R$ and $\langle
T_{\mu\nu} \rangle_R$.  This then allows one to obtain a
self-consistent solution to the mean field and mode equations
(\ref{eq:equationphi2}) and (\ref{eq:equationpsi2}) in a background RW
spacetime (in the test field approximation) or to these equations plus
the semiclassical backreaction equations in a RW spacetime.  But
before getting involved in the backreaction problem (and the
difficulties this presents), it would be very useful to have a way of
estimating how big or small the quantum corrections are beyond test
field approximation. In this section we discuss this issue and present
such an approximation.

For a free quantum scalar field in a general RW spacetime, one of us
has already shown that there is a way of defining a certain
approximate energy-momentum tensor that is covariantly
conserved~\cite{paul99}. In the present paper we extend this method to
interacting quantum fields, and discuss its advantages and
limitations.

The analytic approximations result from a revision of the method of
adiabatic regularization that simplifies the calculations in a RW
spacetime with compact spatial sections, and in the process yields
approximations for the quantities $\langle \psi^2 \rangle_R$ and
$\langle T_{\mu\nu} \rangle_R$.  These analytic approximations give
information about vacuum polarization effects but not particle
production, since particle production is a nonlocal phenomenon.
However they also make it possible to solve the mean field, mode, and
semiclassical backreaction equations in an approximate manner, which
goes beyond the test field approximation, and does not involve a full
backreaction calculation.  This can be useful, for example, if one
wishes to determine under what conditions vacuum polarization effects
will be important and what influence they may have on the mean field
and the spacetime geometry.  In particular, we believe that these
analytic approximations may provide important information for
reheating calculations, just before particle production from the
inflaton field takes place. The analytic approximations provide a
natural way to estimate the change in the vacuum polarization energy
of the inflaton field. During the inflationary regime the mean field
$\phi$ dominates the energy density of the universe. It is reasonable
to expect that when the vacuum polarization energy is of the order of
the mean field energy density, the inflaton field will switch from the
slow-roll regime to the oscillatory behavior, that will eventually
lead to particle production. In this paper we restrict ourselves to
presenting the analytic approximations and our method for obtaining
them.  In future work we will present the applications to reheating.

In order to derive the analytic approximation, we first improve on the
method of adiabatic regularization by expanding the renormalization
counterterms in inverse powers of $k$, keeping only terms which are
ultraviolet divergent.  For the case of compact spatial sections
($\kappa=+1$) the integral is also changed into a sum.  We call the
resulting expressions $\langle \psi^2 \rangle_d$ and $\langle
T_{\mu\nu}\rangle_d$, respectively. In a general RW spacetime they
have the form
\begin{mathletters}
\begin{eqnarray}
\langle \psi^2 \rangle_{d} 
&=& \frac{1}{4 \pi^2 a^2}\int {\mathrm{d}}\mu(k)\;  \frac{1}{k}  
 - \frac{1}{4 \pi^2 a^2}
\int {\mathrm{d}}\bar{\mu}(k) \; \frac{1}{k^3} \left[
\frac{m^2 a^2}{2} + \frac{\lambda 
  \phi^2 a^2}{4}  + \left(\xi - \frac{1}{6} \right)
 \frac{a^2 R}{2} \right]
\; ,
\label{eq:psi-d}
 \\
\langle{T^\eta}_\eta\rangle_{d} &=& 
\frac{1}{4 \pi^2 a^4}\int {\mathrm{d}}\mu(k) 
               \left(k + \frac{1}{k} \left[\frac{m^2 a^2}{2} 
+ \frac{\lambda \phi^2 a^2}{4} 
             - 3 \left(\xi - \frac{1}{6}\right) 
\left(\frac{a'^{\,2}}{a^2} 
             -  \kappa \right)\right] \right) \nonumber \\
                &  &  +\, \frac{1}{4 \pi^2 a^4} 
\int {\mathrm{d}}\bar{\mu}(k) \; \frac{1}{k^3} 
                     \left[ - \frac{m^4 a^4}{8} 
- \frac{m^2 \lambda \phi^2 a^4}{8}
             - \frac{\lambda^2 \phi^4 a^4}{32} 
             - \frac{3 m^2 a^2}{2} \left(\frac{a'^{\,2}}{a^2} 
+ \kappa \right) \right. \nonumber \\
         & & \left.  - \; 6 \left(\xi - \frac{1}{6} \right) 
\frac{\lambda}{8} (\phi^2 a'^{\, 2}
           + 2 \phi \phi' a a' + \kappa \phi^2 a^2)
 +  \left(\xi - \frac{1}{6}\right)^2\, 
                        {^{(1)}H^\eta}_\eta \,\frac{a^4}{4} \right]
\; ,
\label{eq:t00-d}
\\
\langle{T}\rangle_{d} &=& \frac{1}{4 \pi^2 a^4} 
\int {\mathrm{d}} \mu(k) \; \frac{1}{k} \,
  \left\{
m^2 a^2 + \frac{\lambda \phi^2 a^2}{2} 
          -  6\left(\xi - \frac{1}{6}\right) \left(\frac{a''}{a}
            - \frac{a'^{\,2}}{a^2} \right) \right] \nonumber \\  
      &  & +\, \frac{1}{4 \pi^2 a^4} \int {\mathrm{d}} \bar{\mu}(k) \; 
\frac{1}{k^3} 
 \left[- \frac{m^4 a^4}{2} - \frac{m^2 \lambda \phi^2 a^4}{2} 
 - \frac{\lambda^2 \phi^4 a^4}{8} 
        \right. \nonumber \\
 & & \left. - \left(\xi - \frac{1}{6}\right) 
\left[ 3 m^2 a^2 \left(\frac{a''}{a} 
      + \kappa\right) + \frac{\lambda}{4} 
(6 \phi \phi'' a^2 + 6 \phi'^{\,2} a^2 
  + 12 \phi \phi' a a' + 6 \phi^2 a a'' 
+ 6 \kappa \phi^2 a^2) \right] \right. \nonumber \\
   &  & \left.  + \left(\xi - \frac{1}{6}\right)^2\, 
                   {^{(1)}H} \,\frac{a^4}{4} \right\}
\; , 
\label{eq:t-d}
\end{eqnarray}
\end{mathletters} 
with
\begin{eqnarray*}
\begin{array}{lllll}
  \int {\mathrm{d}}\bar{\mu}(k) &\equiv& \int_\varepsilon^{+ \infty }
{\mathrm{d}} k\;  k^2 \;\;\;
&{\rm if}& \kappa = 0, -1 \; ,
\vspace{0.2cm}  
    \nonumber \\
                     &\equiv& \sum_{k = 1}^{+ \infty} k^2 \;\;\; 
&{\rm if}& \kappa = +1 \; .
\end{array}
\end{eqnarray*}
Here $\varepsilon$ is an arbitrary lower limit cutoff and
\begin{mathletters}
\begin{equation}
 {^{(1)}H_{\mu\nu}} = 2 R_{;\mu\nu} - 2 g_{\mu\nu} \Box R - \frac{1}{2}
     g_{\mu\nu} R^2 + 2 R R_{\mu\nu}\; .
\end{equation}
In a RW spacetime it has the components
\begin{eqnarray}
 {^{(1)}H^\eta}_\eta
 &=& - \frac{36 a' a'''}{a^6} + \frac{72 a'^{\,2} a''}{a^7}
  + \frac{18 a''^{\,2}}{a^6} + \frac{36 \kappa
 a'^{\,2}}{a^6} - \frac{18 \kappa^2}{a^4} \; ,
     \label{eq:h100} \\
 {^{(1)}H} &=& - \frac{36 a''''}{a^5} + \frac{144 a' a'''}{a^6}
  - \frac{216 a'^{\,2} a''}{a^7} + \frac{108 a''^{\,2}}{a^6} 
   + \frac{72 \kappa a''}{a^5} - \frac{72 \kappa a'^{\,2}}{a^6} 
\;. \label{eq:h1tr}
\end{eqnarray}
\end{mathletters}
The renormalized energy-momentum tensor is then computed by
subtracting and adding the quantity $\langle T_{\mu\nu}\rangle_{d}$ to
Eq. (\ref{eq:tmnad}), with the result that
\begin{mathletters}
\begin{eqnarray}
\langle T_{\mu\nu} \rangle_{r} 
&\equiv& \langle T_{\mu\nu}\rangle_{n} + \langle T_{\mu\nu}\rangle_{an}  
\; , 
\label{eq:Tmnr}
\\
\langle T_{\mu\nu}\rangle_{n} 
&\equiv& \langle T_{\mu\nu}\rangle_{u} - \langle T_{\mu\nu}\rangle_{d}  
\label{eq:Tmnn}
\; , \\
\langle T_{\mu\nu}\rangle_{an} 
&\equiv& \langle T_{\mu\nu}\rangle_{d} - \langle T_{\mu\nu}\rangle_{ad} 
\label{eq:Tmnan}
\; . 
\end{eqnarray}
\end{mathletters}
In general $ \langle T_{\mu\nu}\rangle_n$ must be computed numerically
while $\langle T_{\mu\nu}\rangle_{an}$ can always be computed
analytically.  The result is
\begin{mathletters}
\begin{eqnarray}
 \langle \psi^2 \rangle_{an} &=&   
 \langle \psi^2 \rangle_{an}^F-
\frac{\lambda \phi^2}{16 \pi^2}
\left\{
1 - \left [
\log \left( \frac{2 \varepsilon }{a \mu}\right)
- \frac{\kappa (\kappa+1)}{2}(\log \varepsilon + \gamma)
\right]
\right\}
\; ,
\label{eq:psi-an}
\\
 \langle {T^\eta}_\eta\rangle_{an} &=& 
\langle {T^\eta}_\eta\rangle_{an}^{ F} 
+ 
\frac{1}{4 \pi^2}
\left\{
-\frac{\lambda \phi^2}{48} \frac{\kappa (\kappa+1)}{2 a^2}
- \frac{\lambda \phi^2}{16} \left(m^2 + \frac{\lambda \phi^2}{2} \right)
+ \frac{\lambda \phi^2}{8} \left(m^2 + \frac{\lambda \phi^2}{4} \right)
\left[
\log \left( \frac{2 \varepsilon }{a \mu}\right)
- \frac{\kappa (\kappa+1)}{2}(\log \varepsilon + \gamma)
\right]
\right\}
\nonumber \\
&+&
\frac{(6 \xi -1) \lambda}{16 \pi^2 a^4}
\left(a^2 \kappa \frac{\phi^2}{2}
+ a'^{\, 2} \frac{\phi^2}{2} + a a' \phi \phi'
\right)
\left[
\log \left( \frac{2 \varepsilon }{a \mu}\right)
- \frac{\kappa (\kappa+1)}{2}(\log \varepsilon + \gamma)
\right]
\nonumber \\
&-&\frac{\lambda}{96 \pi^2 a^4}
\left( a'^{\, 2} \frac{\phi^2}{2} + a a'  \phi \phi'
\right)
-\frac{(6 \xi -1)\lambda}{16 \pi^2 a^4}
\left(a^2 \kappa \frac{ \phi^2}{2}
+ a'^{\, 2} {\phi^2} + a a' \phi \phi'
\right)
\; ,
\label{eq:t00-an}
\\
 \langle T \rangle_{an} &=& \langle T \rangle_{an}^{F} 
+ 
\frac{1}{4 \pi^2}
\left\{
-\frac{\lambda \phi^2}{48} \frac{\kappa (\kappa+1)}{a^2}
- \frac{\lambda \phi^2}{4} \left(\frac{3m^2}{2} 
+ \frac{\lambda \phi^2}{2} \right)
+ \frac{\lambda \phi^2}{4} \left(2m^2 + \frac{\lambda \phi^2}{2} \right)
\left[
\log \left( \frac{2 \varepsilon }{a \mu}\right)
- \frac{\kappa (\kappa+1)}{2}(\log \varepsilon + \gamma)
\right]
\right\}
\nonumber \\
&+&
\frac{(6 \xi -1)\lambda}{16 \pi^2 a^3}
\left( a \kappa { \phi^2}
+ a'' {\phi^2} +2   a'  \phi \phi' 
+ a \phi'^{\, 2}+a \phi \phi''
\right)
\left[
\log \left( \frac{2 \varepsilon }{a \mu}\right)
- \frac{\kappa (\kappa+1)}{2}(\log \varepsilon + \gamma)
\right]
\nonumber \\
&-&\frac{\lambda}{96 \pi^2 a^4}
\left(
3 a^4 \lambda \frac{\phi^4}{4}
+a^2 \phi'^{\, 2}+a^2 \phi \phi''
+a a'' {\phi^2} +2  a a' \phi \phi'
\right)
\nonumber \\
&-&\frac{(6 \xi -1)\lambda}{16 \pi^2 a^4}
\left(3 a^2 \kappa \frac{\phi^2}{2}
+ a'^{\, 2} \frac{\phi^2}{2} +4  a a'  \phi \phi'
+ 2  a a'' \phi^2+a^2 \phi'^2+ a^2 \phi \phi''
\right)
\; ,
\label{eq:t-an}
\end{eqnarray}
\end{mathletters}
with
\begin{eqnarray}
\langle \psi^2 \rangle_{an}^F
&=&
-\frac{a''}{48 \pi^2 a^3}
+ \frac{1}{4 \pi^2}
\left\{
- \frac{m^4}{4}
-\frac{\kappa(\kappa + 1)}{24 a^2}
+ \frac{m^2}{2}
\left[
\log \left( \frac{2 \varepsilon }{a \mu}\right)
- \frac{\kappa (\kappa+1)}{2}(\log \varepsilon + \gamma)
\right]
\right\}
\nonumber \\
&-&
\frac{(\xi - 1/6)R}{8 \pi^2}
\left\{1-
\left[
\log \left( \frac{2 \varepsilon }{a \mu}\right)
- \frac{\kappa (\kappa+1)}{2}(\log \varepsilon + \gamma)
\right]
\right\}
\; ,
\\
\langle {T^\eta}_\eta\rangle_{an}^{F}
&=&
 \frac{1}{2880 \pi^2} \left[
  - \frac{1}{6}\, {^{(1)}{H^\eta}_\eta} + {^{(3)}{H^\eta}_\eta} 
- \frac{3 \kappa (\kappa -1)}{a^4}
        \right] + \frac{m^2}{288 \pi^2} {G^\eta}_\eta
- \frac{m^2 \kappa (\kappa -1)}{192 \pi^2 a^2} 
\nonumber \\ 
  & &   
         - \frac{m^4}{64 \pi^2} \left[\frac{1}{2} 
+ 2\log\left(\frac{\mu a} 
           {2 \varepsilon}\right) 
+ \kappa (\kappa +1)(\gamma+\log \varepsilon )\right] 
      + \left(\xi - \frac{1}{6}\right)  
           \left[ \frac{{^{(1)}{H^\eta}_\eta}}{288 \pi^2} + 
            \frac{\kappa (\kappa -1)}{32 \pi^2 a^4} 
\left(1 + \frac{a'^{\,2}}{a^2} \right) 
     \right. \nonumber \\
         &  & \left. + \frac{m^2}{16 \pi^2} {G^\eta}_\eta
 \left(3 + 2\log\left(
           \frac{\mu a}{2 \varepsilon}\right) + 
         \kappa (\kappa +1) (\gamma 
+ \log \varepsilon )\right) + 
          \frac{3 \kappa  m^2}{8 \pi^2 a^2} \right]  \nonumber \\
     &  &  + \left(\xi - \frac{1}{6}\right)^2  
               \left[ \frac{{^{(1)}{H^\eta}}_\eta}{32 \pi^2} \left(2 + 
         2 \log\left(\frac{\mu a}{2\varepsilon}\right) + 
          \kappa (\kappa +1) ( \gamma
 + \log \varepsilon )\right) 
- \frac{9}{4 \pi^2} \left(
    \frac{a'^{\,2} a''}{a^7} 
+ \frac{\kappa  a'^{\,2}}{a^6} \right)\right]
\; ,
\\
\langle T\rangle_{an}^{F}
&=&
 \frac{1}{2880 \pi^2} \left[
   -\frac{1}{6} \,{^{(1)}}{H} + {^{(3)}}{H} \right] 
    - \frac{m^2}{288 \pi^2} G 
    - \frac{m^2 \kappa (\kappa -1)}{96 \pi^2 a^2} 
   - \frac{m^4}{16 \pi^2} \left[1 +2 \log\left(\frac{\mu a}
           {2 \varepsilon}\right) +  \kappa (\kappa +1)
           ( \gamma + \log \varepsilon)\right] \nonumber \\ 
   &  &    + \left(\xi - \frac{1}{6}\right)  
           \left[ \frac{{^{(1)}{H}}}{288 \pi^2} + 
           \frac{\kappa (\kappa -1)}{16 \pi^2 a^4} \left(\frac{a''}{a} 
- \frac{a'^{\,2}}{a^2} \right) 
- \frac{m^2}{16 \pi^2} G
           \left(3 + 2\log\left(\frac{\mu a}{2 \varepsilon}\right)
         + \kappa (\kappa +1)(\gamma + \log \varepsilon) 
\right) \right. \nonumber \\  
 &  & \left. + \frac{3 \kappa  m^2}{8 \pi^2 a^2} 
- \frac{3 m^2 a'^{\,2}}{8 \pi^2 a^4} 
          \right] 
+ \left(\xi - \frac{1}{6}\right)^2 \left[ \frac{^{(1)}{H}}
         {32 \pi^2} \left(2 + 2 \log\left(\frac{\mu a}{2 \varepsilon}
\right)
          + \kappa (\kappa +1)( \gamma
+ \log \varepsilon)\right) \right. \nonumber \\
  &  &  \left.  - \frac{9}{8 \pi^2} \left(
    \frac{4 a' a'''}{a^6} - \frac{10 a'^{\,2}a''}{a^7} 
    + \frac{3 a''^{\,2}}{a^6} + \frac{4 \kappa  a''}{a^5} 
- \frac{6 \kappa  a'^{\,2}}{a^6} 
    +\frac{\kappa ^2}{a^4} \right)\right] 
\; .
\end{eqnarray}
Here $G_{\mu\nu}$ is the Einstein tensor with components
\begin{mathletters}
\begin{eqnarray}
 {G^\eta}_\eta &=& - \frac{3 a'^{\,2}}{a^4} - \frac{3 \kappa }{a^2} 
\; ,\\
 G &=& -\frac{6 a''}{a^3} - \frac{6 \kappa }{a^2} = -R
\; ,
\end{eqnarray}
\end{mathletters}
and ${^{(3)}H_{\mu\nu}}$ is the tensor
\begin{mathletters}
\begin{equation}
 {^{(3)}H_{\mu\nu}} = {R_\mu}^\rho R_{\rho\nu} - \frac{2}{3} R R_{\mu\nu}
           - \frac{1}{2} R_{\rho\sigma} R^{\rho\sigma} g_{\mu\nu} 
           + \frac{1}{4} R^2 g_{\mu\nu} 
\; ,
\end{equation}
with components
\begin{eqnarray}
{^{(3)}H^\eta}_\eta &=&
 \frac{3 a'^{\,4}}{a^8} + \frac{6 \kappa  a'^{\,2}}{a^6}
                   + \frac{3 \kappa ^2}{a^4}
\; ,  \\
{^{(3)}H} &=& 
\frac{12  a'^{\,2}a''}{a^7} - \frac{12 a'^{\,4}}{a^8}
+ \frac{12 \kappa  a''}{a^5} - \frac{12 \kappa  a'^{\,2}}{a^6} 
\; .
\end{eqnarray}
\end{mathletters}
For a massive field $\mu = m$, while for a massless field $\mu$ is an
arbitrary mass scale.  However, in the massless case each of the terms
containing $\log \mu$ has as a coefficient a multiple of the tensor
${^{(1)}H_{\mu\nu}}$, which comes from an $R^2$ term in the
gravitational Lagrangian.  Thus the terms containing $\log \mu$ simply
correspond to a finite renormalization of the coefficient of the $R^2$
term in the gravitational action.

Note that if $\kappa = +1$, $\langle \psi^2 \rangle_d$ and $\langle
T_{\mu\nu} \rangle_{d}$ contain a sum over $k $, while $\langle \psi^2
\rangle_{ad}$ and $\langle T_{\mu\nu}\rangle_{ad}$ contain an integral
over $k$.  Thus, either the integral must be converted to a sum or the
sum to an integral.  We have converted the sum to an integral using
the Plana sum formula~\cite{Pl1,Pl2,Pl3,Pl4}. This formula is
\begin{equation}
 \sum_{n=m}^{+ \infty} f(n) = \frac{1}{2} \; f(m) 
+ \int_m^{+ \infty} {\mathrm{d}} x \; f(x) 
       + i \int_0^{+ \infty} \frac{{\mathrm{d}} t}{e^{2 \pi t} - 1}
\;  [f(m+it) - f(m-it)] 
\;.
\end{equation}
Because of the way $\langle T_{\mu\nu}\rangle_{d}$ is defined, the
third term in the Plana sum formula can be computed exactly.  In the
traditional form of adiabatic regularization one would convert the
integral in the adiabatic counterterms to a sum using the Plana sum
formula and then substitute the result into Eq. (\ref{eq:tmnad}).
However, if this is done then, for a massive field, it is not possible
to compute the third term in the Plana sum formula analytically.  Thus
the computation of the renormalized energy-momentum tensor is
simplified by our method in the $\kappa=+1$ case.  Clearly the same
simplification would occur if one was using compact spatial sections
for $\kappa=0$ or $\kappa=-1$ RW spacetimes.

\section{Conservation of the energy-momentum tensor}
\label{sec:conservation}

In this section we show that the renormalized energy-momentum tensor
for the $\lambda \phi^4$ theory (at one-loop) in a RW
spacetime is covariantly conserved.  In a RW spacetime there is only
one nontrivial conservation equation which is
\begin{equation}
{\langle {T^\eta}_{\eta}\rangle}^{C}_{R \, ,\eta} 
+  \frac{4 a'}{a}{\langle {T^\eta}_{\eta}\rangle}^{C}_{R} 
 - \frac{a'}{a} {\langle {T}\rangle}^{C}_{R}  = 0
\; .
\label{eq:cons}
\end{equation}

We have shown that the energy-momentum tensor for the field can be
divided into a classical and a quantum part.  First consider the
classical part which in a RW spacetime has the components
(\ref{eq:t00-b}) and (\ref{eq:t-b}).  Substituting into
Eq. (\ref{eq:cons}) and using (\ref{eq:phi-b}) we find
\begin{equation}
{T^\eta}_{\eta ,\eta} 
+  \frac{4 a'}{a}{T^\eta}_\eta - \frac{a'}{a} T = 
- \frac{\lambda}{2} \phi \phi' 
\langle \psi^2 \rangle_R \;.
\end{equation}
Thus the classical energy-momentum tensor is conserved only if there
is no quantum correction to the effective mass of the mean field.

The energy-momentum tensor for the quantum part consists of the
difference between the unrenormalized part and the adiabatic
counterterms.  The components of the unrenormalized part are given in
Eqs. (\ref{eq:t00u}) and (\ref{eq:tru}).  If they are substituted into
Eq. (\ref{eq:cons}) and the mode equation (\ref{eq:modef}) is used
then one finds
\begin{equation}
\langle {T^\eta}_{\eta}\rangle_{u \, ,\eta} 
+  \frac{4 a'}{a} \langle {T^\eta}_\eta\rangle_u - \frac{a'}{a} 
\langle T\rangle_u 
 =  \frac{\lambda}{2} \phi \phi' \langle \psi^2 \rangle_u \;.
\end{equation}
If one substitutes the adiabatic counterterms (\ref{eq:t00-ad}) and
(\ref{eq:t-ad}) into Eq. (\ref{eq:cons}), and compares the result with
Eq. (\ref{eq:psiad}) one finds
\begin{equation}
\langle {T^\eta}_{\eta}\rangle_{ad \, ,\eta} 
+  \frac{4 a'}{a} \langle {T^\eta}_\eta\rangle_{ad} 
- \frac{a'}{a} \langle T\rangle_{ad} 
 =  \frac{\lambda}{2} \phi \phi' \langle \psi^2 \rangle_{ad} \;.
\end{equation}
Combining these results and using (\ref{eq:psiad}) and
(\ref{eq:tmnad}), one finds that the total renormalized
energy-momentum tensor, classical plus quantum, is conserved.

One further finds that if the analytic approximation is used in place
of $\langle\psi^2\rangle_R$ in the equation for the mean field, and if
it is used for the quantum energy-momentum tensor, then the analytic
approximate energy-momentum tensor is conserved.  This means that one
can use the analytic approximation to define a consistent set of
equations for the mean and the quantum fields, and to be the source in
the right hand-side of Einstein's equation to solve a first
approximation to the backreaction problem.

Thus $\langle \psi^2 \rangle_{an}$ and $\langle
T_{\mu\nu}\rangle_{an}$ can be used in place of $\langle \psi^2
\rangle_{R}$ and $\langle T_{\mu\nu}\rangle_{R}$ in Eqs.\
(\ref{eq:equationphi2}), (\ref{eq:tensorC}), and (\ref{eq:tensorQ}) to
obtain an analytic approximation for the system.  For $\kappa = 0, -1$
RW spacetimes the terms being approximated contain the arbitrary
constant $\varepsilon$.  This means the approximation is not unique
unless the coefficients of the $\log \varepsilon$ terms vanish.  It is
important to note that this is only true when using $\langle \psi^2
\rangle_{an}$ and $\langle T_{\mu\nu}\rangle_{an}$ as an analytic
approximation.  The $\varepsilon$ dependent terms do not appear in the
fully renormalized expressions of these quantities.

{From} the DeWitt-Schwinger expansion~\cite{Christensen} it is known
that for a free quantum field in the large mass limit $\langle \psi^2
\rangle_{R}$ and $\langle T_{\mu\nu}\rangle_{R}$ have leading order
terms proportional to $1/m^2$.  Thus, the analytic approximations for
these quantities are not good approximations in this limit.  Previous
numerical work~\cite{A34} indicates that the relevant condition is
likely to be $m a\ll 1$. The analytically approximated quantities are
also local in the sense that they depend on the scale factor and its
derivatives at a given time $\eta$.  Therefore, they cannot accurately
describe particle production effects which are inherently nonlocal.
However for massless fields they should allow one to estimate how
important vacuum polarization effects are, and how they qualitatively
effect the evolution of the system.

\section{Summary}

We have used adiabatic regularization to renormalize a scalar field
theory with a quartic self-coupling of the form $\lambda \phi^4$ in an
arbitrary RW spacetime.  We have found that the energy-momentum tensor
can be naturally split into two parts, a ``classical'' contribution
(which corresponds to the energy-momentum tensor of a classical scalar
field with a quartic interaction in a RW spacetime) and a ``quantum''
piece (which corresponds to the energy-momentum tensor of a free
quantum scalar field with the time dependent mass $m^2 + \frac{\lambda
\phi^2}{2}$).  We have displayed the renormalization counterterms for
both the energy-momentum tensor and the contribution of the quantum
fluctuations to the effective mass of the mean field at one-loop
order.  We have directly checked to see if the energy momentum tensor
is covariantly conserved and found that while the entire tensor is
conserved, its classical and quantum contributions are not separately
conserved.

By using a variant on the adiabatic regularization method we have
derived analytic approximations for the energy-momentum tensor and the
contribution of the quantum fluctuations to the effective mass of the
mean field.  We have shown that the approximate energy-momentum tensor
is covariantly conserved.  Thus the analytic approximations can be
used in a self-consistent way to find approximate solutions to the
mean field and backreaction equations.  The approximations can provide
a useful tool for learning about vacuum polarization effects for
massless fields.  However, they are not useful for massive fields in
the large mass limit.  They do not give any significant amount of
information about particle production, which is a nonlocal phenomenon.

The approximations could be of use in reheating calculations, in
particular the transition from the slow-roll dynamics of the inflaton
field to the oscillatory behavior around the minimum of the potential,
(which is believed to produce particles of lighter masses), and in the
context of relativistic heavy ion collision as a way of estimating the
physical energy density and pressure of the vacuum and thermal
excitations. The advantage of this approximation is that one obtains
analytic expressions for the quantum piece of the energy-momentum
tensor, without need to solve exactly the mode equation (which is the
most difficult part to implement in numeric computations). In this
way, it is relatively easy to study the backreaction problem of the
full system (mean field, quantum fluctuations, and gravitational
field).

We believe that this program can be carried forward, and improved
easily.  We plan to extend the present approach and approximations to
two-loop order, and to the $1/N$ expansion. We also plan to perform
some specific calculations with applications to reheating and early
density perturbations.

\section{Acknowledgements}

The authors wish to express their gratitude to Salman Habib and Emil
Mottola for their incisive comments on the manuscript and endless
hours of fruitful discussion.  Part of this work was done while P.\
R.\ A.\ was visiting at Montana State University.  He would like to
thank W.\ Hiscock and the entire Department of Physics there for their
hospitality.  This work was supported in part by grant number
Phy-9800971 from the National Science Foundation. C.\ M.-P. was
partially supported by the Department of Energy under contract
W-7405-ENG-36.

\end{document}